\documentclass{aip-cp}

\usepackage[numbers]{natbib}
\usepackage{rotating}
\usepackage{graphicx}

\def\be{\begin{equation}}
\def\ee{\end{equation}}
\def\bea{\begin{eqnarray}}
\def\eea{\end{eqnarray}}

\def\gev{\, {\rm GeV}}
\def\mev{\, {\rm MeV}}

\newcommand{\sigmaSD}{\sigma_{\rm SD}}
\newcommand{\gsim}{\lower.7ex\hbox{$\;\stackrel{\textstyle>}{\sim}\;$}}
\newcommand{\lsim}{\lower.7ex\hbox{$\;\stackrel{\textstyle<}{\sim}\;$}}

\newcommand{\pb}{\rm pb}

\newcommand{\cm}{\rm cm}
\newcommand{\km}{\rm km}
\newcommand{\s}{\rm s}

\begin{document}

\title{Monoenergetic Neutrinos From Dark Matter Annihilation: Issues of Exposure }

\author[aff1]{Jason Kumar\corref{cor1}}

\affil[aff1]{Department of Physics and Astronomy, University of Hawai'i,
  Honolulu, HI 96822, USA}
\corresp[cor1]{Corresponding author: jkumar@hawaii.edu}

\maketitle

\begin{abstract}
We consider searches for dark matter annihilation in the Sun resulting in  monoenergetic
neutrinos, produced either directly or through the decay of stopped pions and kaons.  We find that this
strategy is very successful at increasing the signal-to-background ratio, but that current
experiments may be signal limited.  We discuss the exposures need to fully exploit this
search strategy.
\end{abstract}

\section{INTRODUCTION}

If dark matter exhibits spin-dependent scattering with nuclei, then a key detection strategy
is the search for neutrinos arising from dark matter
annihilation in the core of the Sun.  Most such searches look for a smooth excess of
charged lepton events above the expected background, rather than for a monoenergetic
signal.  The reason is two-fold: the theoretical models which have received the most
consideration produce a continuum of neutrinos rather than monoenergetic neutrinos, and,
in any case, many
detection strategies are not able to fully reconstruct
the neutrino energy.

We will see that both of the arguments above can be evaded.  For models in which either dark matter is
not a Majorana fermion or in which flavor symmetry violation is non-minimal, dark matter annihilation
in the Sun can produce light Standard Model fermions, allowing significant branching fractions for the
processes $\bar X X \rightarrow \bar \nu \nu, \bar u u, \bar d d, \bar s s$.  Direct annihilation to neutrinos
produces a high-energy monoenergetic neutrino/anti-neutrino signal, whereas annihilation to light quarks results in
hadronic cascades that produce a multitude of stopped $\pi^+$ and $K^+$ in the Sun.  The decays of these
stopped $\pi^+/K^+$ produce a low-energy monoenergetic neutrino signal~\cite{PionKaon}. If these monoenergetic (anti-)neutrinos
oscillate to $\bar \nu_e / \nu_e$, then a charged-current (CC) interaction in the detector will produce an
electromagnetic cascade whose energy is easily contained within the detector, allowing one to resolve
the neutrino line~\cite{Learned:2009rv,Kumar:2009ws,Kumar:2011hi}.

A search for a neutrino line signal will benefit from a large signal-to-background ratio~\cite{Kumar:2015nja}.
The main limitation of this search strategy lies in obtaining an exposure large enough to benefit from the
enhanced signal-to-background ratio.  In this proceedings contribution, we will first describe two classes of
models in which dark matter annihilation in the Sun produces monoenergetic neutrinos.  We will then describe
the sensitivities of liquid scintillation (LS), liquid argon time projection chamber
(LArTPC), and water Cherenkov (WC) detectors to a neutrino line signal.  We discuss signal limitations and optimal exposures for these
types of experiments, and conclude with a discussion of our results.

\section{MODELS AND DETECTION STRATEGIES}

For most commonly considered theoretical models of dark matter, annihilation to light fermion/anti-fermion pairs is either
chirality or $p$-wave suppressed.  The reason for this suppression can be understood from considerations of angular
momentum (see, for example,~\cite{Kumar:2013iva}).  If dark matter is a Majorana fermion, then the initial state wavefunction
must be totally anti-symmetric, implying that $s$-wave annihilation ($XX \rightarrow \bar f f$) can only occur from a $J=0$ initial state.
But if the $\bar f f$ final state has $J=0$, then $f$ and $\bar f$ must have the same helicity, and thus arise from different Weyl spinors.
If one assumes minimal flavor violation, then any matrix element which mixes those Weyl spinors must be proportional to $m_f$,
resulting in the chirality-suppression of the $s$-wave annihilation matrix element to light fermions.

However, this argument has
assumed minimal flavor violation, and that dark matter is a Majorana fermion.  Although these assumptions are valid in
constrained scenarios such as the CMSSM, dark matter need not be Majorana in general, and flavor violation may be non-minimal even in
a generic MSSM model.  If either of these assumptions are relaxed, then dark matter annihilation to light fermion/anti-fermion
pairs could have an ${\cal O}(1)$ branching fraction.

There is another reason why dark matter annihilation in the Sun to light quarks is typically ignored: the hadronization of light
quarks typically produces light hadrons which stop in the Sun before decaying.  These decays thus produce a very soft neutrino spectrum,
whereas typical searches for dark matter annihilation in the Sun focus on energetic neutrinos.  But, as was pointed out in~\cite{Rott:2012qb,Bernal:2012qh},
the stopping of energetic hadrons in a dense medium produces a large number of mesons such as $\pi^+$ and
$K^+$, whose decays also produce neutrinos.  Essentially one is trading a hard neutrino flux for a softer flux, but with a larger amplitude.  But a key point relevant here is
that the decay of $\pi^+$ and $K^+$ produce monoenergetic neutrinos through the processes $\pi^+, K^+ \rightarrow \nu_\mu \, \mu^+$
(the $\pi^-/K^-$ produced by hadronic cascades are Coulomb-captured by nuclei).

We thus consider two classes of models which produce monoenergetic (anti-)neutrinos:
\begin{itemize}
\item{{\it Direct annihilation to $\bar \nu \nu$:} produces monoenergetic $\bar \nu \nu$ pairs with $E_{\bar \nu , \nu} = m_X > \gev$.}
\item{{\it Annihilation to $\bar q q$, $q=u,d,s$:} produces monoenergetic $\nu_\mu$ with $E_{\nu_\mu} \sim 30\mev \, (236\mev)$ from the
decay of stopped $\pi^+$ ($K^+$) with branching fraction $\sim 100\%$ ($\sim 64\%$).  }
\end{itemize}

Typical searches for neutrinos arising from dark matter annihilation in the Sun focus on $\bar \nu_\mu, \nu_\mu$, but for a monoenergetic neutrino
search, one should focus on neutrinos which have oscillated into $\bar \nu_e$,  $\nu_e$~\cite{Kumar:2009ws,Kumar:2011hi}, because electron (anti-)neutrinos will produce
an $e^\pm$ after a charged-current interaction in the detector.  The result will be a relatively short-range cascade, whose energy will be
entirely contained within the detector, thus permitting reconstruction of the original energy of the incoming neutrino.

\section{SENSITIVITIES}

The four quantities which enter into the calculation of detector sensitivity are the flux of monoenergetic
neutrinos arising from dark matter annihilation in the Sun, the atmospheric neutrino background flux,
the effective area of the detector, and the energy resolution.

\subsection{Neutrino flux from dark matter annihilation}

Dark matter particles are gravitationally captured and collect in the core of the Sun after scattering
against solar nuclei.    The rate of dark matter capture ($\Gamma_C$) can be written as
$\Gamma_C = C_0^{SD} (m_X) \times \sigmaSD^p \times [(\rho_X /\rho_\odot)(\bar v / 270~\km/\s)]^{-1}$~\cite{Gould:1987ir},
where $\rho_X$ is the dark matter density, $\rho_\odot = 0.3~\gev / \cm^3$, $\sigmaSD^p$ is the dark matter-proton
spin-dependent elastic scattering cross section, and $\bar v$ is the dark matter velocity dispersion if one assumes a
Maxwell-Boltzmann distribution.   Values of $C_0^{SD} (m_X)$ can be found,
for example, in~\cite{Gao:2011bq,Kumar:2012uh}.  Note, we assume spin-dependent scattering because dark matter-nucleon spin-independent
scattering is already very tightly constrained by direct detection experiments.

If the Sun is in equilibrium, then the rate of dark matter capture is twice the rate
of dark matter annihilation.  In this limit, the annihilation rate is completely
determined by $m_X$ and $\sigmaSD^p$ (for a particular dark matter distribution).
The flux of monoenergetic electron (anti-)neutrinos at the detector arising from dark matter
annihilation in the Sun is in turn determined by the annihilation rate, the final state channel,
and the effects of neutrino oscillations (including matter effects).

If dark matter annihilates directly to monoenergetic neutrinos, then each annihilation produces
a neutrino and an anti-neutrino with $E_\nu = m_X/2$.  We will assume that dark matter annihilates to
each neutrino flavor with equal probability; in this case, after including the effects of neutrino
oscillations, the neutrino flux at a detector on Earth will be flavor-independent~\cite{Lehnert:2007fv}.
If dark matter annihilates to light quarks, then the number of monoenergetic $\nu_\mu$ produced
is determined by $n_{\pi^+,K^+} (m_X)$, the number of stopped $\pi^+$ and $K^+$ which arise from each annihilation.
We determine the $n_{\pi^+,K^+} (m_X)$ by simulating
the process $XX \rightarrow \bar q q$, including the effects of showering and hadronization, using Pythia 8.2~\cite{Sjostrand:2014zea}.
The interactions of the annihilation products with the nuclear medium in the Sun are simulated using
GEANT~\cite{Agostinelli:2002hh}.  Finally, the fraction of the injected monoenergetic $\nu_\mu$ which will oscillate to $\nu_e$ by
the time they reach the detector, $F_{\nu_e}$, can be determined from Reference~\cite{Lehnert:2007fv}.  We will assume a normal
hierarchy, in which case $F_{\nu_e} (E_\nu = 30~\mev)\sim 0.36$,  $F_{\nu_e} (E_\nu = 236~\mev)\sim 0.46$.

\subsection{Backgrounds}

For the energetic neutrinos produced by direct dark matter annihilation, the dominant background comes
from atmospheric electron (anti-)neutrinos.  For $E_\nu > 1~\gev$, the angle-averaged flux can be estimated as~\cite{Honda:2011nf}
\bea
d^2 \Phi_{atm}^{\nu_e} / d\Omega dE_\nu  &\sim&  (4.17 \times 10^{-2} \cm^{-2} \s^{-1} {\rm sr}^{-1} \gev^{-1} )
\times \left(0.80 + E_\nu / \gev \right)^{-3.490},
\nonumber\\
d^2 \Phi_{atm}^{\bar \nu_e} / d\Omega dE_\nu &\sim&  (2.42 \times 10^{-2} \cm^{-2} \s^{-1} {\rm sr}^{-1} \gev^{-1} )
\times \left(0.53 + E_\nu / \gev \right)^{-3.417}.
\eea

For monoenergetic neutrinos arising from stopped $\pi^+$, there can be other sources of background as well.
Atmospheric $\nu_\mu$ can produce a low-energy $\mu^-$ which does not produce a Cherenkov cone, but decays to a low-energy
$e^-$.  This background will be difficult to distinguish at a water Cherenkov detector (and is larger than that arising from
atmospheric $\nu_e$ by a factor $\sim 10$), but we will assume that it can
be substantially reduced by track reconstruction at a LS or LArTPC detector.  Monoenergetic neutrinos can also arise from
stopped cosmic ray pions, but this background is small compared to that from atmospheric neutrinos~\cite{PionKaon}.  The relevant fluxes
are approximately given by~\cite{Battistoni:2005pd}:
\bea
(d^2 \Phi_{atm}^{\nu_e} / d\Omega dE_\nu  ) [E_\nu =30~\mev ] &\sim& 1~\cm^{-2} \s^{-1} {\rm sr}^{-1} \gev^{-1} ,
\nonumber\\
(d^2 \Phi_{atm}^{\nu_e} / d\Omega dE_\nu  ) [E_\nu =236~\mev ] &\sim& 0.1~\cm^{-2} \s^{-1} {\rm sr}^{-1} \gev^{-1} .
\eea
For scattering off C or O, $\sim 15~\mev$ of the neutrino energy is lost to the change in nucleus
binding energy; thus for $30~\mev$ $\nu_e$s we focus only on argon.
Although stopped $\pi^+/K^+$ decay produces only monoenergetic $\nu$, an atmospheric $\bar \nu$ would be difficult
to distinguish; conservatively, we assume that this produces a factor $\times 2$ increase in background.

\subsection{Effective area}

The effective area of the detector may be expressed as $A_{eff} = \sigma \times N_T$, where $\sigma$ is the neutrino-nucleus
charged-current scattering cross section and $N_T$ is the number of target nuclei in the fiducial volume (chosen such that
the produced electromagnetic shower will be fully contained).  For $E_\nu \gg 1~\gev$, the charged-current scattering cross
section is dominated by deep inelastic scattering, and the neutrino-nucleon scattering cross section may be written as~\cite{Edsjo:1997hp}
\bea
\sigma_{\nu N} \sim (6.66 \times 10^{-3}~\pb) (E_\nu /GeV) , \qquad
\sigma_{\bar \nu N} \sim (3.25 \times 10^{-3}~\pb) (E_\nu /GeV) .
\eea
At low energies, however, the neutrino scatters off the entire nucleus.
The total scattering cross sections for the relevant nuclei can be estimated using GENIE~\cite{Andreopoulos:2009rq}
and Reference~\cite{Kolbe:2003ys} (at $E_\nu=30~\mev$), yielding:
\bea
\sigma_{\rm Ar} (E_\nu =236~\mev) &\sim& 5.2 \times 10^{-2}~\pb , \qquad \sigma_{\rm Ar} (E_\nu =30~\mev) \sim 1.8 \times 10^{-4}~\pb ,
\nonumber\\
\sigma_{\rm C} (E_\nu =236~\mev) &\sim& 1.6 \times 10^{-2}~\pb ,
\nonumber\\
\sigma_{\rm water} (E_\nu =236~\mev) &\sim& 2.0 \times 10^{-2}~\pb .
\eea

Energy resolutions ($\epsilon$) on the order of a few percent are possible for LS~\cite{Peltoniemi:2009xx} and LArTPC~\cite{Adams:2013qkq} detectors,
which are very efficient at energy collection.  For WC detectors, the energy resolution
is typically worse.  We will thus only consider WC detectors for $236~\mev$ neutrinos, which lose little energy in hadronic
showers.
High energy neutrinos from dark matter annihilation also tend to produce charged leptons which
are concentrated within an rms cone of the Sun parameterized by $\theta = 0.37 \sqrt{10~\gev / E_\nu }$; a search for leptons within
this cone will also significantly reduce background.  But for the low-energy neutrinos produced by stopped $\pi^+$, $K^+$ decay,
the charged leptons are produced nearly isotropically.

\begin{figure}[t]
  \begin{tabular}{c}
    \includegraphics[width=0.45\textwidth]{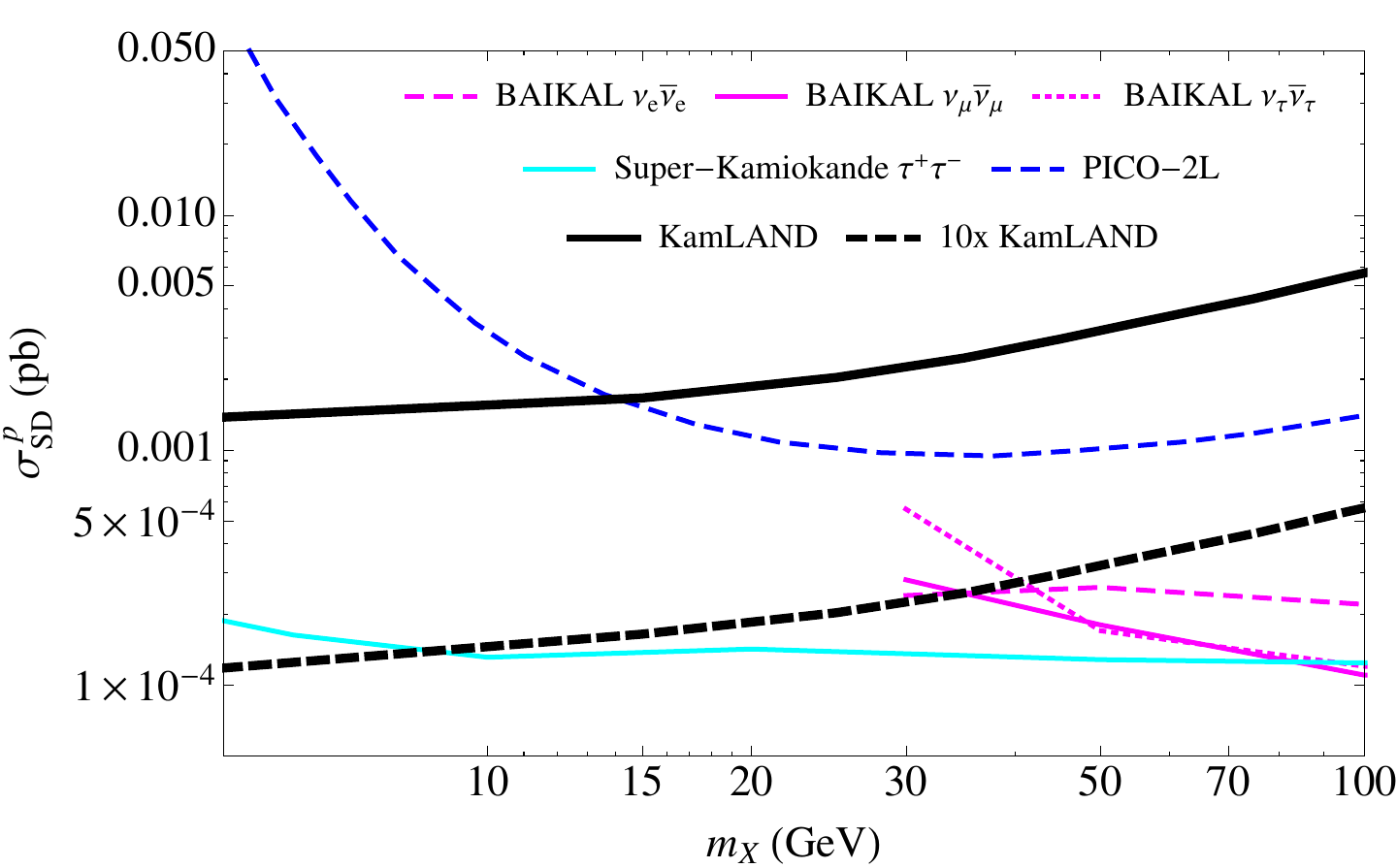}
    \includegraphics[width=0.41\textwidth]{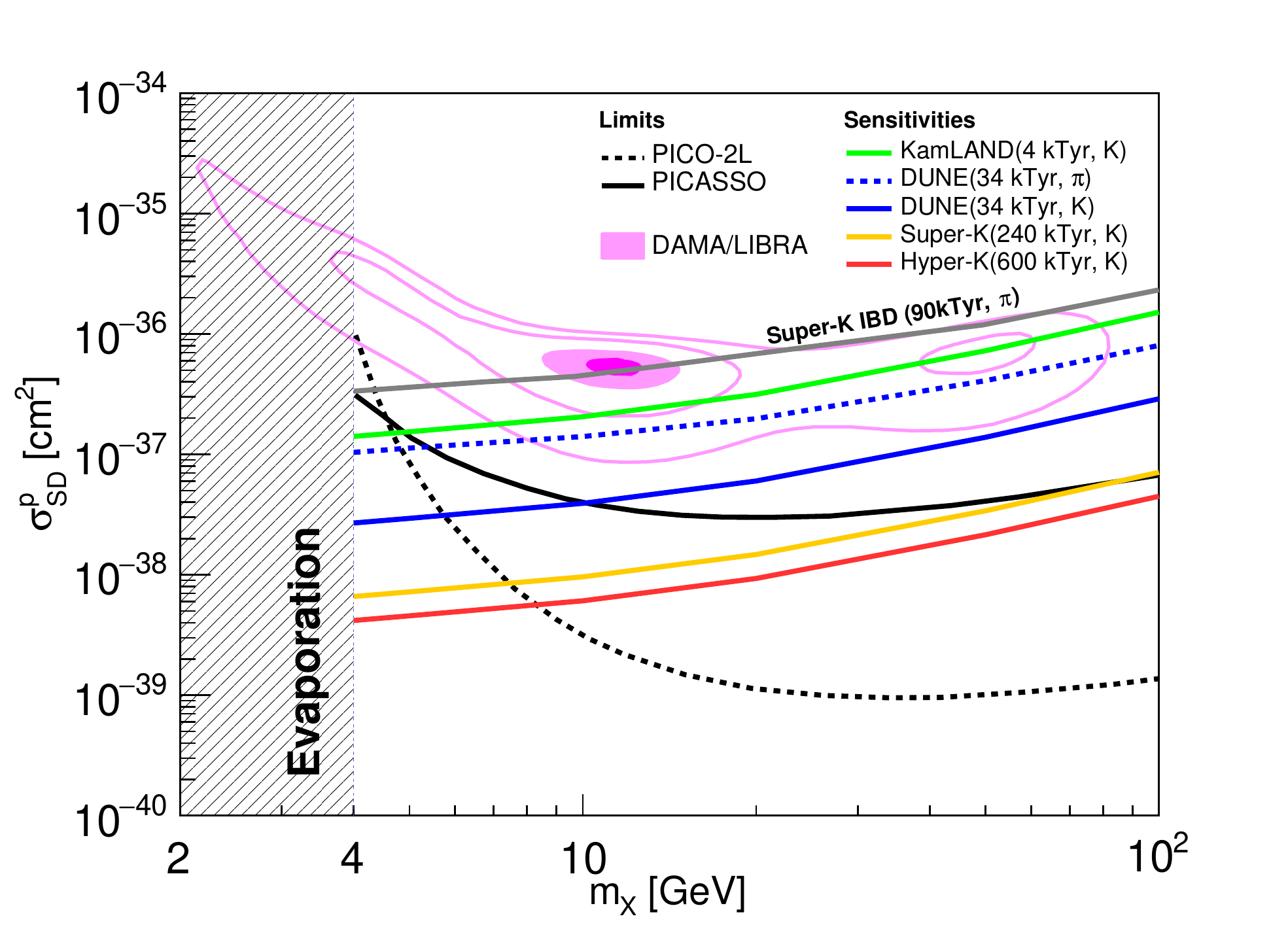}
  \end{tabular}
    \caption{90\% CL upper limit on $\sigmaSD^p$ as a function of dark matter mass, assuming
    $XX \rightarrow \bar \nu \nu$ (left) and $XX \rightarrow \bar u u$ (right).  Figure is
    described in text. (Figures courtesy of Pearl Sandick and Seongjin In.)
\label{fig:limitplot}}
\end{figure}

In Figure~\ref{fig:limitplot} we plot the 90\%CL sensitivity to $\sigmaSD^p$ which arises from a search for dark matter
annihilation entirely through the process $XX \rightarrow \bar \nu \nu$ (flavor-independent)~\cite{Kumar:2015nja}(left panel)
and from a search for dark matter annihilation entirely to through the process $XX \rightarrow \bar u u$ (the
$\bar d d$ channel is identical)~\cite{PionKaon}(right panel).  We assume the number of observed events is set by the
expected background.
Our benchmark detectors are KamLAND (4 kT yr LS), a benchmark LS detector (40 kT yr LS),
DUNE (34 kT yr LArTPC), Super-K (240 kT yr WC)
and Hyper-K (600 kT yr WC).
As reasonable benchmark values, we will take the energy resolution of all detectors to be $\epsilon= 10\%$ for the
$236~\mev$ neutrinos produced by $K^+$ decay.  For the $30~\mev$ neutrinos produced by $\pi^+$ decay, we will only
consider DUNE (also with $\epsilon = 10\%$), since the binding energies for C and O are significant compared to
$30~\mev$.  For the high energy $\nu$s produced by direct annihilation, we consider KamLAND ($\epsilon = 5\%$) and
a benchmark LS detector with a 40 kT yr exposure ($\epsilon = 3\%$).
We also show upper limits from Baikal NT200 detector~\cite{Avrorin:2014swy} (assuming annihilation to $\bar \nu_e \nu_e$,
$\bar \nu_\mu \nu_\mu$, $\bar \nu_\tau \nu_\tau$), Super-K (assuming annihilation to $\tau^+ \tau^-$~\cite{Choi:2015ara}),
the Super-K (90 kT yr) stopped $\pi^+$ analysis of~\cite{Rott:2012qb,Bernal:2012qh} and
from PICASSO~\cite{Archambault:2012pm} and PICO-2L~\cite{Amole:2015lsj}, as labeled.
Also plotted are the 90\% CL, $3$, $5$, and $7\sigma$ signal regions (from innermost to outermost) for DAMA/LIBRA~\cite{Savage:2008er}.
Details of the analysis are given in Refs.~\cite{PionKaon,Kumar:2015nja}.

\section{EXPOSURE}

As expected, both direct annihilation to neutrinos and annihilation to light quarks result in an
enhanced signal-to-background ratio ($S/B$), as compared to more typical continuum searches.  In fact, the
most significant limitation to these searches arises from a lack of exposure.  For example, if dark
matter annihilates directly to neutrinos, then KamLAND is likely to have far less than one event
the relevant energy bin~\cite{Kumar:2015nja}.  Its sensitivity is thus controlled entirely by necessity of producing even a few
signal events, given its exposure.  On the other hand, KamLAND has no useful sensitivity to the stopped
$\pi^+$ channel; although the background is expected to be less than one event over KamLAND's current exposure,
any models which could produce a few signal events over that exposure would already have been easily ruled
out by direct detection experiments.

In order to fully exploit this search strategy, one should be sensitive to models for which
$S/B \sim 1$; if not, then sensitivity will continue to grow linearly
with increased exposure.  To be sensitive to a dark matter model, one must have an exposure large enough to
expect at least a few signal events.  As a rough criterion, we thus ask two questions:
\begin{itemize}
\item{What is the {\it optimal exposure} needed to obtain one expected background event ($B \sim 1$)?}
\item{For that exposure, what $\sigmaSD^p$ would yield one expected signal event ($S \sim B \sim 1$)?  This is roughly
the {\it optimal sensitivity}.}
\end{itemize}

$S/B$ does not depend on the target material or the exposure but only on the signal and background fluxes.  For a fixed channel,
$S$ is determined only by the capture rate and by the number of monoenergetic neutrinos per annihilation, yielding $S \propto \sigmaSD^p$.
Since $B \propto \epsilon$, the optimal sensitivity is proportional to $\epsilon$, while the optimal exposure is proportional to $\epsilon^{-1}$.
For each channel, the optimal sensitivities and LArTPC exposures are listed in Table~\ref{Tab:Exposure} for $m_X =10~\gev$.

We see that the $K^+$ channel optimal sensitivity suffers from the small number of $K^+$ produced per annihilation, compared
to $\pi^+$ channel.  But the increase in $A_{eff}$ at higher energy implies that the $K^+$ channel will yield greater
sensitivity for small exposure.  At the optimal exposure of the $\pi^+$ channel, the $\pi^+$ and $K^+$ channels will
have comparable sensitivity, and subsequently the sensitivity of both channels with grow as the square root of exposure.
The $\bar \nu \nu$  channel
benefits from a smaller background and larger effective area, provided this annihilation channel is available.

\begin{center}
\begin{table}
\begin{tabular}{|c|c|c|c|}
                \hline
                channel & optimal sensitivity ($\pb$) & optimal exposure (kT yr) & optimal for \\
                \hline
                $\pi^+$ & $0.21~\epsilon$ & $17 \epsilon^{-1}$ & $S/B$ $(\bar q q)$ \\
                $K^+$ & $1.2~\epsilon$ & $0.07 \epsilon^{-1}$ & rapid exploitation $(\bar q q)$\\
                $\bar \nu \nu$ & $1.7 \times 10^{-4}~\epsilon$ & $12 \epsilon^{-1}$ & sensitivity ($\bar \nu \nu$) \\
                \hline
\end{tabular}
\caption{Optimal sensitivity and optimal LArTPC exposure for each channel, for $m_X = 10~\gev$.}
\label{Tab:Exposure}
\end{table}
\end{center}

\section{CONCLUSION}

Searches for dark matter annihilation in the Sun typically look for a smooth excess of events
above background.  But if dark matter is not a Majorana fermion, or if flavor violation is not
minimal, then dark matter annihilation can easily produce monoenergetic neutrinos.
We have considered two such classes of models with distinct energy ranges: models in the which dark matter
directly annihilates to $\bar \nu \nu$ pairs with $E_\nu >~\gev$, and models where dark matter annihilation
produced stopped $\pi^+$ and $K^+$, whose decays produce monoenergetic neutrinos ($E_\nu = 30~\mev$ or $236~\mev$).
The energy of the neutrino can then be resolved if it oscillates to $\nu_e$ and interacts in the detector via
a CC interaction.

This type of search strategy results in a significant reduction in background.  The difficulty
lies in obtaining a large enough exposure to fully exploit the strategy, which is largely signal limited.  KamLAND
has a very long runtime, but still operates in the regime where it is signal limited; unless a much larger LS detector
can be built, it is difficult to make progress with that technology.  DUNE (34kT LArTPC) can fully exploit the kaon channel
within a few months, but would need closer to 5 years to fully exploit the direct production or pion channels, despite the
much larger $S/B$ for those channels.
Super-Kamiokande is a water Cherenkov detector with both large size and large runtime; it has the exposure to fully exploit
the $\pi^+$ and $K^+$ channels with current data, but has much larger backgrounds due to the lower energy resolutions inherent in a
WC detector, potentially weakening the usefulness of this strategy.

One general result is clear: future detectors with larger fiducial volumes would
greatly enhance the usefulness of this search strategy.

\section{ACKNOWLEDGMENTS}
We are grateful to Seongjin In, Carsten Rott, Pearl Sandick and David Yaylali
for collaboration.  We are grateful to John G.~Learned and Michinari Sakai for useful discussions.
This research is funded in part by NSF CAREER grant PHY-1250573.
We thank CETUP* (Center for Theoretical Underground Physics and Related Areas),
for its hospitality and partial support during the 2015 Summer Program.


\nocite{*}
\bibliographystyle{aipnum-cp}%
\bibliography{PPC}%

\end{document}